# Nano-Imaging of Chiro-Optical Force

AUTHOR NAMES

Junsuke Yamanishi,[*,1] Hyo-Yong Ahn,[1,2] and Hiromi Okamoto[*,1]

AUTHOR ADDRESS

[1]*Institute for Molecular Science, National Institutes of Natural Sciences, 38 Nishigonaka, Myodaiji, Okazaki, Aichi 444-8585, Japan*

[2]*Center for Novel Science Initiatives, National Institutes of Natural Sciences, 4-3-13 Toranomon, Minato-ku, Tokyo 105-0001, Japan*

ABSTRACT

Nanoscopic observation of chiro-optical phenomena is essential in wide scientific areas but has measurement difficulties; hence, its physics are still unknown. Currently, in most cases, chiro-optical phenomena have been investigated by polarized light handling far-field measurements or via predictions by theoretical simulations. To obtain a full understanding of the physics of chiro-optical systems and derive the full potentials, it is essential to perform in situ observation of the chiro-optical effect from the individual parts because the macroscopic chiro-optical effect cannot be translated directly into microscopic effects. In the present study, we observed the chiro-optical responses at the nanoscale level by detecting the chiro-optical forces, which were generated by illumination of the material/probe system with circularly polarized light. The induced optical force was dependent on the handedness of the incident circularly polarized light and well correlated to the electromagnetically simulated differential intensity of the longitudinal electric field. Our results facilitate the clarification of chiro-optical phenomena at the nanoscale level and could innovate chiro-optical nanotechnologies.

One of the essential characteristics of chiral structures is the chiro-optical effect; this effect shows different responses to left-circularly polarized (LCP) and right-circularly polarized (RCP) light [1–3]. To analyze the chirality of matter, macroscopic (ensemble) measurements of polarization-handling spectroscopy methods, characterized by circular dichroism (CD) or optical rotation (OR), are widely used. The chiro-optical effect is generally weak in spectroscopic signals, which causes difficulty to measure the nanoscopic physics at the single molecule or nanostructure level. One of the ways to overcome the limitation of this inherent weakness of chiro-optical signals is to utilize enhanced chiral interactions between molecules or nanomaterials and chiral optical fields through superchiral fields [4, 5]. In particular, plasmonic structures have been recognized as representative materials that generate superchiral fields and enhance the optical activity of systems [3, 6–8]. To fully understand and utilize the full potential of chiro-optical systems, local in situ observation of the effects from individual parts is essential, as the macroscopic (ensemble) chiro-optical effect is not straightforward for the analysis of microscopic phenomena. However, nanoscopic technologies for chiro-optical measurements are presently limited [9–11], and the chiro-optical phenomena at the nanoscale level have not been effectively determined; conventionally, these phenomena have been investigated through indirect information from macroscopic far-field optical measurements and/or predictions by theoretical simulations [7, 8, 12].

For a new nanoscopic chiro-optical measurement, utilization of optical force is promising [13–15]. In a previous study, a plasmonic tip with a spiral (chiral) structure was used to detect the field induced by circularly polarized (CP) light, where the tip moved depending on the handedness of incident CP light [14]. The study targeted the detection of the CP-light-induced force on a chiral tip, and the technology was difficult to routinely utilize for the investigation of nanoscale chiro-optical effects of materials in their original form. For optical force measurements based on a scanning probe, another sophisticated technique, photoinduced force microscopy (PiFM), has been developed and extensively investigated [16–20]. In PiFM, both the tip and the sample are irradiated by polarized incident light, and the nanoscopic optical responses of matter are observed via the motion of the tip induced by the incident light. There are two types of detection modes currently used in PiFM, which are classified by the detected force. One of the modes to measure the optical force is based essentially on the dipole–dipole force [17, 19–23], and the other measures the photoinduced thermal expansion of the samples or tips [18, 24, 25]. In this paper, we call the former the optical force mode (OF-mode) and the latter the thermal mode to distinguish the two. Theoretical studies have shown that PiFM, particularly in OF-mode, has the potential to detect chiro-optical phenomena at the nanoscale level [26]. Achievement of chiro-optical PiFM would provide not only an experimental breakthrough in nanoscale physics and chemistry of chirality but also major technological advances for pharmaceutical and biological sciences. In the present study, we aimed to image the chiro-optical forces confined at the nanoscale level based on OF-mode PiFM.

For the detection of the chiro-optical effect on the optical force in PiFM, precise optical force detection is essential because the chiro-optical effect (i.e., the difference between the optical forces with LCP and RCP light) is generally small relative to the total strength of the optical force. In this regard, the signal arising from the photothermal effects needs to be eliminated in OF-mode observation because it degrades the optical force detection. The heterodyne frequency modulation (HFM) technique needs to be performed in OF-mode PiFM for the selective detection of the optical force to eliminate the photothermal effect of the tip and the sample [19, 27]. We utilized the HFM technique with frequency modulation atomic force microscopy (FM-AFM) for the ambient PiFM to perform noncontact tip scanning on the sample surface, as illustrated in Fig. 1a. FM-AFM proves that he probe is in a noncontact region that facilitates the measurement of the optical force. If the probe is in contact with the sample surface, the detected signal is mainly caused by the photothermal expansion of the probe and the sample [18, 24, 25]. The gold-coated tip and the sample were illuminated from the bottom side of the sample substrate by CP light from a diode laser with a wavelength of 785 nm. Right-handed gammadion gold nanostructures, which are sometimes used as the specimens for chiro-optical effect observation [7, 28, 29], were fabricated on a quartz substrate with electron beam lithography and lift off technique, as shown in the model illustration (Fig. 1b) and the scanning electron microscopy (SEM) image (Fig. 1c). The SEM image of the left-handed gold gammadion nanostructures is shown in Fig. S1 in the Supporting Information. The CD spectra of the gammadion structures are plotted in Fig. 1c, where the wavelength of 785 nm is indicated by the dashed line. The black solid and red dotted curves represent the CD spectra of the left- and right-handed gammadion structures, respectively. The extinction spectra of the structures are shown in Fig. S2 in the Supporting Information. The wavelength of 785 nm is outside the CD spectral bands, and apparently, the nanostructures are chiro-optically not very active. However, from the electromagnetic simulation results (Fig. S7 in Supporting Information), the induced chiral electric near fields are found to strongly appear on the top (probe tip side) of the nanostructures, and thus, we used this wavelength for the nanoscopic chiro-optical PiFM measurements.

A topographic (FM-AFM) image of the gammadion structures is shown in Fig. 2a. The structures in the FM-AFM image had vertically (in the *z*-direction) rounded edges rather than the structures with sharp edges as in the SEM image (Fig. 1c). These rounded edges occurred because the tip was scanning in the long-range force region on the surface where the force between the tip and the sample had a long-range decay length [30]; thus, as a consequence, the topography was observed as a smeared image. An observed PiFM image is shown in Fig. 2b; the structure was illuminated with RCP light. The photoinduced force signal in the HFM is given as $\Delta f(f_m)X$ [19, 27]. Here, the frequency shift signal in FM-AFM ($\Delta f$) is modulated at $f_m$ (i.e., $\Delta f(f_m)$), and the in-phase signal to the laser modulation of the signal is lock-in-detected as $\Delta f(f_m)X$ (see Supporting Information for more details). This quantity was shown by the color scale of the image. The repulsive optical force (negative $\Delta f(f_m)X$ value) was found in the entire imaging area, and the repulsive force decreased on the metal nanostructures. This occurred because an attractive optical gradient force (relatively positive $\Delta f(f_m)X$ value) was exhibited on the nanostructures, whereas the scattering force on the probe by the light acted as a repulsive force throughout the imaging region. Note that the PiFM signal on the gammadion structures could also become large due to the lower scattering force caused by less forward-scattered light. Although the PiFM image with RCP light in Fig. 2b appeared to follow the topographic structure (Fig. 2a), the image under illumination of LCP light clearly showed different features from that with RCP light (Fig. 2c). Substantial differences were observed at the edges of the gammadion structures. The LCP light illumination of the gammadion structures generated a strong optical gradient force at the edge positions.

The image features observed are well reproduced by the simulation based on the finite element method (FEM) implemented in the commercial simulation software COMSOL Multiphysics, as described in the following. The model of the simulation is depicted in Fig. 2d. This model effectively reproduces the experimental CD spectra in Fig. 1d, as shown in Fig. S3 in the Supporting Information. In the simulation of the images in Figs. 2e and f, we plotted the intensity of the $z$-component of the electric field ($E_z^2$) at the plane 110 nm from the surface of the quartz substrate. If we model the gold-coated tip as an isotropic gold nanosphere (this model was considered to be reasonable and often used [19, 22]), the optical gradient force exerted on the tip (i.e., the nanosphere in the model) is expressed as $\boldsymbol{F}_{\text{grad}} = \alpha \nabla |\boldsymbol{E}|^2/4$. Here, $\alpha$ denotes the electric polarizability of a gold nanosphere, and $\boldsymbol{E}$ is the electric field at the position of the tip dipole. Based on this theoretical expression of the optical gradient force, the image contrast should have a correlation to the intensity of the electric field intensity on the structure ($|\boldsymbol{E}|^2$). To note, the field induced by the tip dipole was neglected in the discussion above. Specifically, it is necessary to consider the self-consistent electromagnetic interactions, including the tip [22]. In addition, the transverse components of the electric field ($E_x$, $E_y$) are much smaller than the longitudinal component ($E_z$), on the horizontal surface of the metal nanostructure, as shown in Fig. S4. This result is easily understood from Maxwell's equations of the metal-insulator boundary condition. Here, we compare the experimental PiFM images with the simulated maps of $E_z^2$, which are normalized by the intensity of the incident light. The simulated maps of $E_z^2$ are shown in Figs. 2e and f for RCP and LCP illumination, respectively. With LCP illumination, the simulated map also exhibits strong $E_z^2$ at the corner edge positions of the structures in a similar way as that of the PiFM image. Differences from the experimental images are observed at the central parts of the metal nanostructures. This disagreement appears to arise from less scattering force (less forward-scattered light) at the center of the structure, as discussed in the previous paragraph. The OF-mode PiFM signal includes both the optical gradient force and the optical scattering force. It might be thought that this mixing of the forces in the detection causes difficulty in the interpretation of the experimental PiFM results for nanoscopic chiro-optical activity. However, this difficulty can be solved by reducing the oscillating amplitude of the cantilever to a smaller value than the decay length of the scattering force with the further improved force sensitivity because the optical gradient force has a shorter decay length than that of the scattering force [17].

The structures of the experimental PiFM images do not show complete four-fold rotational symmetry, while the actual shapes of the sample nanostructures have nearly four-fold symmetry, as shown in Fig. 1c. This asymmetry might be due to the asymmetric tip shape and/or the tilted angle of the cantilever to the sample surface. The cantilever is usually tilted to the sample surface for the approach. Although the results of the measurement show these nonideal features arising from the experimental conditions, the images are clearly dependent on the handedness of the incident circularly polarized light.

To investigate the correspondence between the experimental (Figs. 2b and c) and simulated (Figs. 2e and f) images, we examined the line profiles near the edges of the gammadion structures (along the white lines indicated in the insets of Figs. 2g and h). Figure 2g represents the profiles for the images with RCP illumination (Figs. 2b and e), and Fig. 2h shows those with LCP illumination (Figs. 2c and f). The line profiles of Figs. 2g and h show that the image features have similar trends between the experimental and simulated results; two humps appear in the profiles, and the left humps are higher for both the experimental and theoretical curves in Fig. 2g under RCP illumination, while the right humps are higher in Fig. 2h under LCP illumination. These characteristic features under CP light oppositely appear for the left-handed gammadion structure (see Fig. S8 in Supporting Information).

To understand the nanoscopic chiro-optical effect, the differential PiFM image between LCP and RCP illumination conditions is essential. The experimental differential image is depicted in Fig. 3a, and the simulated differential map ($\Delta E_z^2 = E_{zL}^2 - E_{zR}^2$) is shown in Fig. 3b. The experimental differential image is well correlated to the corresponding simulated image; specifically, large differential values appear at the corner edges in both images. This observation indicates that this technology is promising for research on nanoscopic chiro-optical effects. However, the differential PiFM image between LCP and RCP illumination is not completely adequate for the discussion of the nanoscopic chiro-optical effect of the optical gradient force because the detected signal includes not only the optical gradient force but also the optical scattering force, and the scattered light is not the same for RCP and LCP illumination. Hence, the contribution of the scattering force will worsen the validity of the observed nanoscopic chiro-optical activity. The reliability of the dissymmetry factor of the gradient force can be even worse since it is calculated from the observed forces where the scattering forces are commingling (see Fig. S9 in Supporting Information). However, the contribution of the optical gradient force becomes dominant when the amplitude of the cantilever is small. The differential of the optical gradient force thus obtained under the small cantilever amplitude, which reflects the electric near-field intensity, is sufficiently appropriate and useful for quantitative discussion of the local chiro-optical properties. As measurement sensitivity improves, these assessments should become feasible in future studies. Therefore, this pioneering technology is promising for analyzing the nanoscopic chiro-optical activity of materials.

Presently, we observed only chiral gold nanostructures and achieved a spatial resolution of several tens of nanometers. However, this technique has the potential to observe chiro-optical effects at even a higher resolution on single- or sub-nm scales [19]. Furthermore, the tip-induced superchiral field may highly improve the detectable limit of the chiro-optical forces, and the technique may enable observation of the chirality of even a single molecule.

In the present study, we observed nanoscopic chiro-optical forces with OF-mode PiFM using the HFM technique. On the right-handed gammadion structure, a strong optical gradient force appeared at the edges of the structure under illumination with LCP light at 785 nm. The left-handed gammadion produced a similar result under RCP illumination. The measured differential image between LCP- and RCP-light illuminations was well correlated to the difference in the electric-field intensity near the nanostructure calculated with FEM. The present measurement technology could provide novel insights into the physics of chiral optical forces and chiro-optical nanoscopy. For this study, we targeted chiral metal nanostructures. However, this technique has the potential to measure the chiral responses of molecules if we can fully take the advantage of the spatial resolution of the PiFM and the enhanced sensitivity owing to the superchiral fields associated with the plasmons on the tip. The present PiFM-based optical measurement is anticipated to be applied to chemical, biological, and pharmaceutical sciences, where the chirality of molecules plays an essential role.

ASSOCIATED CONTENT

**Supporting Information**.

The following files are available free of charge.

Description of the experimental details, SEM image of the left-handed gammadion gold nanoparticle, extinction spectra of the gammadion structures, simulated CD spectra of the gold gammadion nanostructures, simulation of the electric field on the nanostructures, chiro-optical force imaging with different handed nanostructures, and imaging dissymmetrical factor of photoinduced force. (PDF)

AUTHOR INFORMATION

**Corresponding Author**

Junsuke Yamanishi

*Institute for Molecular Science, National Institutes of Natural Sciences, 38 Nishigonaka, Myodaiji, Okazaki, Aichi 444-8585, Japan*

Email: yamanishi@ims.ac.jp

Hiromi Okamoto

*Institute for Molecular Science, National Institutes of Natural Sciences, 38 Nishigonaka, Myodaiji, Okazaki, Aichi 444-8585, Japan*

Email: aho@ims.ac.jp

**Author Contributions**

The manuscript was written through contributions from all authors. All authors have given approval to the final version of the manuscript. J.Y. conceived the nanoscopic measurement of the chiro-optical effect. J.Y. performed the experiments and analyzed the data. J.Y. performed the electromagnetic simulation. J.Y. and H.O. wrote the paper. All authors discussed the results and commented on the manuscript.

**Funding Sources**

This work was supported by JSPS KAKENHI (Grant Numbers JP19K23589, JP20J00160, JP20K15119, JP21H04641, JP21K18884, JP22H01901, and JP22H05135). This work was also supported by the Frontier Photonic Sciences Project of the National Institutes of Natural Sciences (NINS) (Grant Numbers 01212003, 01213002, and 01212201) and by the NINS program for cross-disciplinary study (Grant Number 01312103).

**Notes**

The authors declare no competing financial interest.

ACKNOWLEDGMENT

We are grateful to H. Yamane (Osaka Research Institute of Industrial Science and Technology) and H. Ishihara (Osaka University) for helpful advice on the sample selection. We would like to express our gratitude to the Equipment Development Center of the Institute for Molecular Science for their assistance with the experimental setup.

ABBREVIATIONS

LCP, left-circularly polarized; RCP, right-circularly polarized; CD, circular dichroism; OR, optical rotation; CP, circularly polarized; PiFM, photoinduced force microscopy; OF, optical force; HFM, Heterodyne frequency modulation; FM-AFM, frequency modulation atomic force microscopy; SEM, scanning electron microscopy.

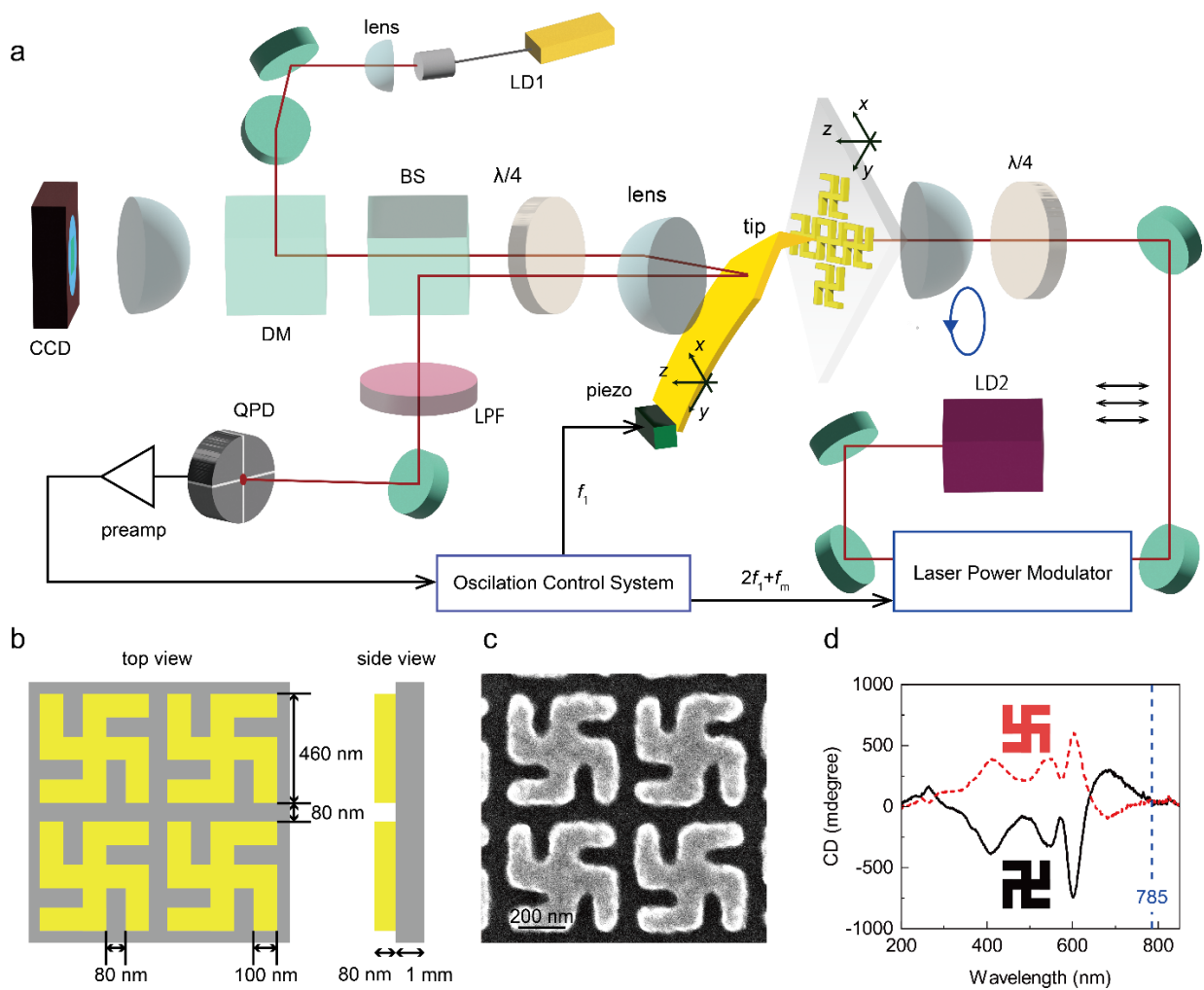

**Figure 1.** Measurement of the chiro-optical force a) Experimental setup for photoinduced force microscopy (PiFM). The following abbreviations are used: LD, laser diode; DM, dichroic mirror; BS, beam splitter; LPF, longpass filter; QPD, quadratic photodiode; Preamp, preamplifier for QPD; λ/4, a quarter waveplate. b) Structural model of the sample (righthanded gammadion structures). c) SEM image of the right-handed gammadion structures. d) CD spectra of the gammadion gold nanostructures. Black and red plots are those of the left- and right-handed nanostructures, respectively. The blue dotted line indicates the wavelength of the incident light (785 nm) used in the PiFM measurements.

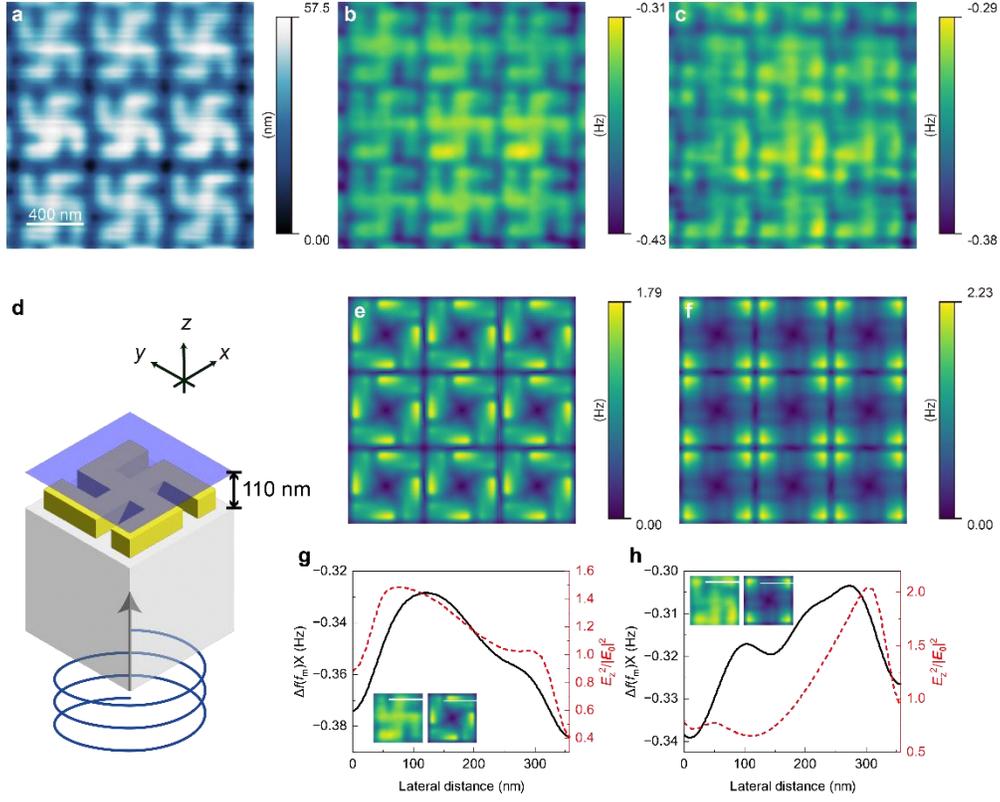

**Figure 2.** Images by photoinduced force microscopy. a) Topographic image (AFM). b, c) PiFM images with RCP- and LCP-light, respectively. $\Delta f = -70$ Hz, $A = 70$ nm, $P_{pp} \sim 30$ mW, and $f_m = 570$ Hz. d) Model for the simulation by FEM. e, f) Simulated results of PiFM images. The images represent the intensity of the longitudinal components of the electric field ($E_z^2$) with RCP- and LCP-light illumination. The image plane is 110 nm away from the surface of the substrates, as shown in Fig. 2d. g, h) Line profiles along the white lines in the insets. Black and red plots are those for the experimental and simulated results, respectively.

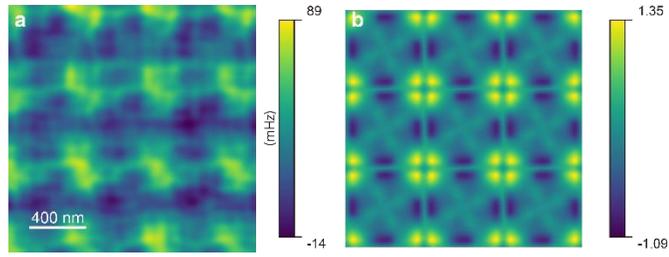

**Figure 3.** Nanoscopic images of the chiro-optical effect obtained via PiFM. a) Differential PiFM image obtained with the experimental results in Figs. 2b and c ($\Delta f(f_m)X_L - \Delta f(f_m)X_R$). b) Differential PiFM image obtained with the simulated results in Figs. 2e and f ($\Delta E_z^2 = E_{zL}^2 - E_{zR}^2$).

# Nano-Imaging of Chiro-Optical Force

Junsuke Yamanishi,[*][1] Hyo-Yong Ahn,[1,2] and Hiromi Okamoto[†][1]

[1]*Institute for Molecular Science, National Institutes of Natural Sciences, 38 Nishigonaka, Myodaiji, Okazaki, Aichi 444-8585, Japan*

[2]*Center for Novel Science Initiatives, National Institutes of Natural Sciences, 4-3-13 Toranomon, Minato-ku, Tokyo 105-0001, Japan*

[*]yamanishi@ims.ac.jp
[†]aho@ims.ac.jp

# Supporting Information

## 1 Experimental Details

The reported measurements were performed using a laboratory-built PiFM apparatus at room temperature in an ambient environment. We used the frequency modulation atomic force microscopy (FM-AFM) mode, in which the tip–sample distance was controlled by the shift of the resonance frequency of the cantilever ($\Delta f$) in non-contact region [1]. We utilized the heterodyne frequency modulation (HFM) technique for the selective detection of the optical force. The modulation frequencies of the laser light ($2f_1 + f_m$) in the HFM technique [2] were set to $2f_1 + 570$ Hz, where $f_1$ is the first resonance frequency of the cantilever ($f_1 = 74.0$ kHz). The laser wavelength was $\lambda_1 = 785$ nm. In the HFM technique, the optical force components have the same phase as the laser modulation. To measure the same phase as the laser modulation, the phase delay due to the electronic circuit used for modulation of the laser power was compensated by measuring the delay. Because of this compensation, a photoinduced force with the same phase as the laser modulation could be detected as a lock-in X component (LIX) in the HFM technique. In this paper, this signal is represented by $\Delta f(f_m)X$. The laser diode was driven such that the laser intensity was $P_{pp} \sim 30$ mW. The cantilever used in the measurements was a gold-coated silicon cantilever (OPUS 240AC-GG, Micromash) with a spring constant of $k \sim 2$ N/m. The Q factor was $\sim 220$. During the measurements, the driven amplitude ($A_1$) in FM-AFM was maintained at 70 nm. A voltage of 10 V was applied on the tip. The images shown in Figs. 2 and 3 in the main text were observed in constant frequency shift feedback mode ($\Delta f = -70$ Hz). The PiFM images were processed by averaging and filtering.

As reported in the main text, the gammadion gold nanostructures on a quartz substrate were observed as the samples. The thickness of the substrate was 1 mm. The gammadion structures were fabricated by electron beam lithography - lift off technique. The circular dichroism (CD) and the extinction spectra of the gold gammadion structures were measured by a commercial circular dichroism spectrometer (J-1500, JASCO).

## 2 Scanning Electron Microscopy Image of the Gammadion Structures

The scanning electron microscopy (SEM) image of the left-handed gammadion gold nanoparticle is shown in Fig. S1. From the SEM images in Fig. 1c in the main text and Fig. S1, we confirmed that the shapes of the structures were well-fabricated as the model in Fig. 1b in the main text.

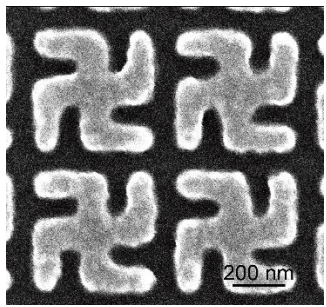

**Figure S1.** A SEM image of the left-handed gammadion structures.

# 3 Extinction Spectra of the Gammadion Structures

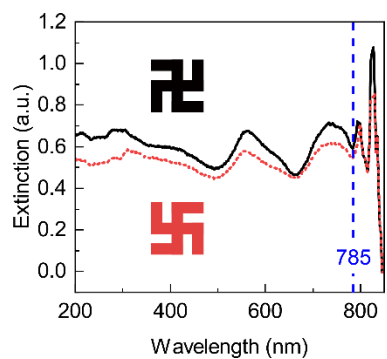

**Figure S2.** Experimental extinction spectra of the fabricated gammadion nanostructures. Black solid and red dotted plots are those of the left- and right-handed nanostructures, respectively. Blue dashed line indicates the wavelength of the incident light (785 nm) used in our PiFM measurements.

# 4 Simulation of the Electric Field on the Nanostructures

We simulated the electromagnetic field using a finite element method (FEM) package software COMSOL Multiphysics. We, then, calculated the CD spectra of the gammadion nanostructures from the simulated results of the absorption cross-section ($C_{abs}$). Figure S3 shows the simulated CD spectra for left- and right-handed gammadions. It should be noted that the experimental CD spectra were obtained by measuring the extinction spectra for left circularly polarized (LCP) and right circularly polarized (RCP) light, which is not directly correlated to $C_{abs}$. However, the simulations of the CD spectra reproduce those of the experimental results in Fig. 1d in the main text at a satisfactory level as shown in Fig. S3. This agreement supports that the simulation reproduces well the chiro-optical effect also in the near-field regime.

In the main text, we discussed the agreement between the experimental PiFM and the simulated images of the gammadion structures (Figs. 2g, h). We chose the electric field intensity at a plane of $z = 110$ nm in the simulation (30 nm above the top surface of the gammadion structures), for the comparison with the experimental results. We chose the plane at $z = 110$ nm for the following reason. In FM-AFM measurement, roughly speaking, the tip–sample force of the middle position of the cantilever vibration (or the force averaged over the amplitude of the cantilever) is detected [3, 4]. In our measurements, the amplitude of the cantilever was 70 nm, hence, the plane of $z = 110$ nm for our comparison with the experiment is reasonable considering that the thickness of the structure was 80 nm (i.e., 80 nm + 70 nm/2 = 115 nm). The validity of the assumption can be confirmed from the simulation results of the electric field intensity at the several planes with different $z$ as represented at Fig. S4. The image planes are at $z = $ 90 nm (a, b), 110 nm (c, d), 130 nm (e, f), 150 nm (g, h), and 200 nm (i, j), respectively. From the comparison of these simulated images, $z = 110$ nm (c, d) seems to show good agreement of the imaging contrasts to the experimental images of PiFM in Figs. 2b and c, as discussed in the main text, in relation to Figs. 2g and h.

A noticeable difference between the experimental PiFM images and the simulated images is found for the image contrast at the center of the gammadion structures. The experimental PiFM images show an attractive force at the center that is similar to the outer parts of the gammadions, whereas the simulation shows weak electric field enhancement, i.e., weak attractive gradient force, at the center. The relatively large attractive forces at the centers in experimental PiFM images compared to that of the simulation might be attributed to the relatively small scattering force at the centers in the experiment, because of the less forward scattering. This is supported by the weak electric field intensity at the centers observed for the simulation in Fig. S4.

It seems that the experimental PiFM images agree with the simulated intensity of $z$ component of the electric field on the gammadion structures particularly at the outer areas, as discussed in the main text (Figs. 2g, h), even though the optical gradient force is proportional to the intensity of the total electromagnetic field (including $x$, $y$, and $z$ components). This is because the $x$ and $y$ components of the electric field are small on the metal nanostructure as shown in Fig. S5, where the simulated transverse ($x$ and $y$) components of the electric field at $z = 110$ nm above the substrate surface are mapped. This result is also understandable from the Maxwell's equations of the metal-insulator boundary condition. However, at the interspaces between the gammadion unit structures at $z = 110$ nm, the simulated intensities of the transverse components of the electric field are not small relative to that of $z$ component, as is visible in Fig. S4. This enhancement does not appear in the experimental PiFM images of Figs. 2b and c in the main text. This discrepancy is considered to be originated in the following fact. In the PiFM measurement, the tip–substrate distance was not constant while the distance was kept constant in the simulation: the tip is in the gaps between the unit gammadion structures. Figure S6 shows simulated distance dependence of the longitudinal and the transverse components of the electric field intensities at three positions under LCP and RCP illumination. In the $z < 80$ nm region in the gaps between the unit gammadion structures (Figs. S6a and b, yellow-shaded area), the electric field intensity under LCP and RCP illumination conditions are small compared with that just above the edge of the gammadion structure ($z \sim 80$ nm) (Fig. S6c, unshaded area; note that the scales of the ordinates are different among a, b, and c). In the $z < 80$ nm region of the Figs. S6a and b, the intensity varies non-monotonically. On the other hand, the intensity right above the edge of the gammadion structure ($z > 80$ nm) in Fig. S6c decreases monotonically with increasing the tip–sample distance. The non-monotonous change of the field intensity in the gaps between the gammadion structures reduces the average gradient force, because the oscillation amplitude of the tip (~70 nm) covers both the increasing and decreasing intensity regions. Another factor of the difference between the experimental and simulated images in the gaps (interspaces) is the strong scattering force at the positions. The experimental PiFM images demonstrated a negatively large force at the interspaces between the metallic gammadion structures, as shown in Figs. 2b and c in the main text, probably because of these multiple factors.

We chose the wavelength of the incident light for the PiFM measurements to be 785 nm. This wavelength is not rigorously resonant with the macroscopic CD band as shown in Fig. 1d in the main text. However, we chose this wavelength because both the intensity of the induced electric field on the top side (i.e., the side where the tip exists) and the intensity difference between LCP and RCP illumination conditions had relatively large magnitudes. When we use other wavelengths, the induced electric field and/or the difference between LCP and RCP illumination conditions are small, as shown Fig. S7 where the longitudinal intensity maps of the electric field at various wavelengths (535, 625, 785, and 855 nm) are shown for LCP and RCP illumination conditions.

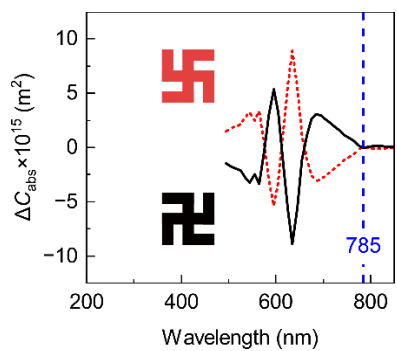

**Figure S3.** Simulated CD spectra of the gold gammadion nanostructures. Black (solid) and red (dashed) plots represent those of the left- and right-handed nanostructures, respectively. Blue dashed line indicates the wavelength of the incident light (785 nm) used in our PiFM measurements.

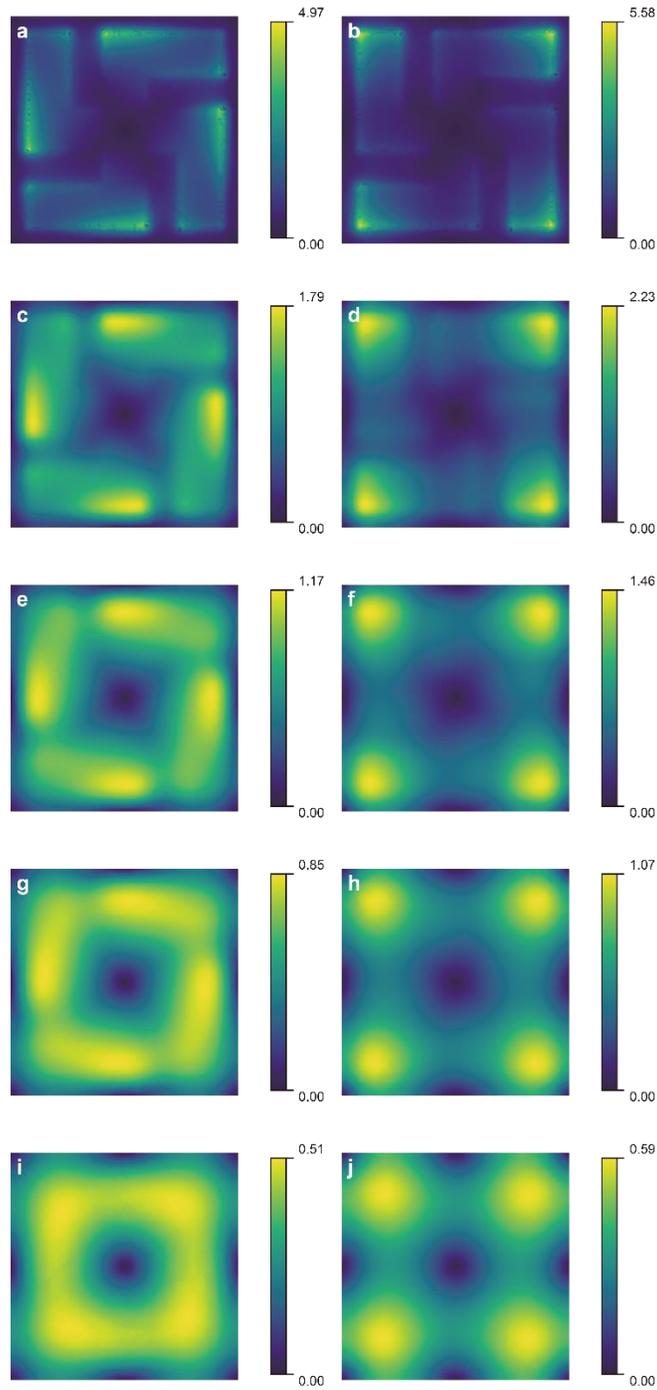

**Figure S4.** Simulated intensity of the longitudinal (*z* component) electric field. Those at *z* = 90 nm (a, b), 110 nm (c, d), 130 nm (e, f), 150 nm (g, h), and 200 nm (i, j), respectively. The incident polarizations are (a, c, e, g, i) RCP and (b, d, f, h, j) LCP.

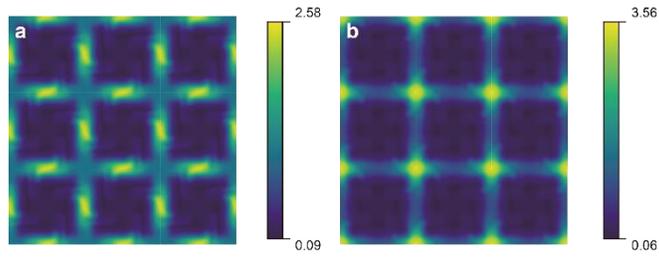

**Figure S5.** Simulated intensity maps of the transverse (i.e., *x* and *y* components) electric field on the gold gammadion nanostructure. Panels a and b show the maps of the field intensities under RCP- and LCP-light illumination, respectively. The image plane is 110 nm from the surface of the substrates, as illustrated in Fig. 2d in the main text.

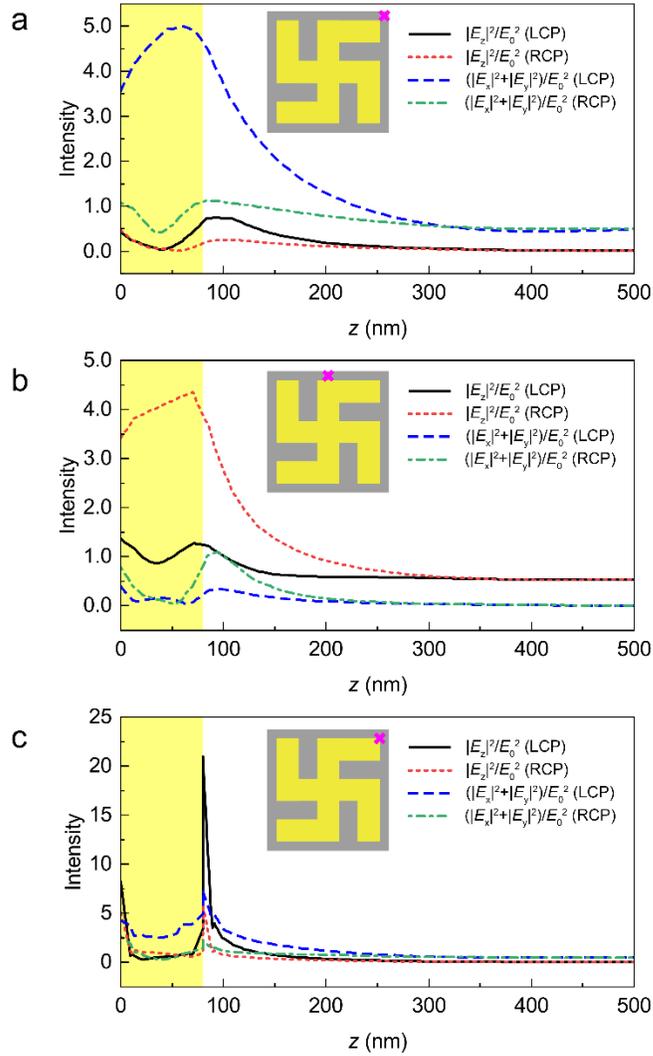

**Figure S6.** Simulated distance (from the substrate surface) dependence of the electric field intensity with LCP and RCP illumination conditions. The positions where the simulations were performed for are indicated in the insets with the x symbols. Yellow shadows represent the height regions where the gold nanostructures are present. Black solid and red dotted plots are the longitudinal intensities of the electric field with LCP and RCP illumination, respectively. Blue dashed and green dot-dashed curves represent transverse intensities of the electric field with LCP and RCP illumination, respectively.

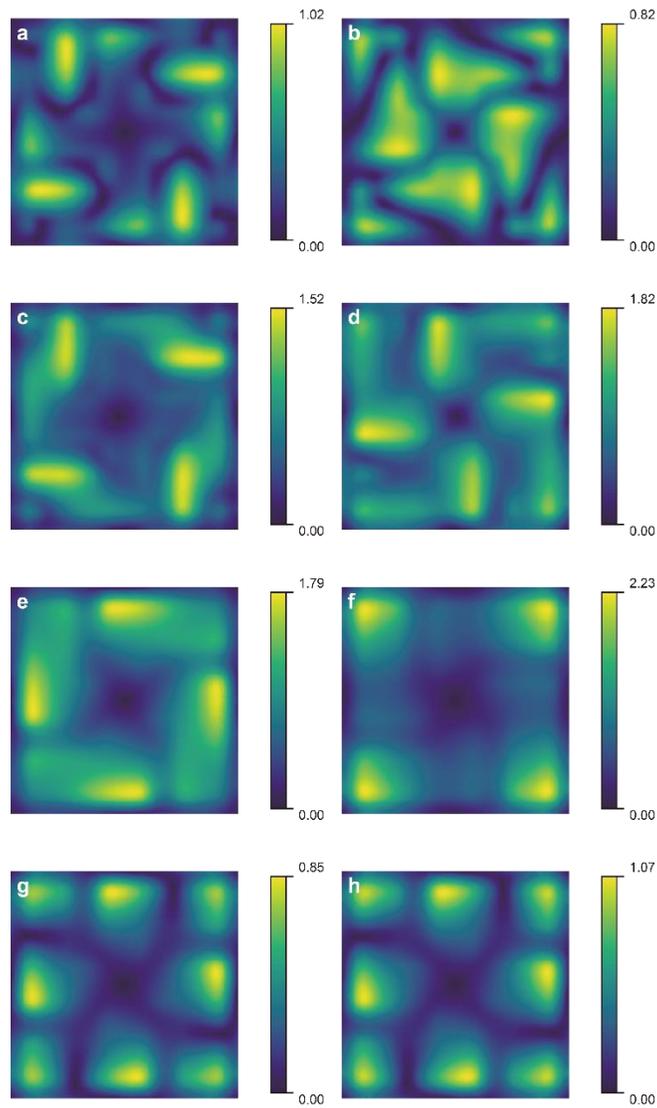

**Figure S7.** Wavelength dependence of the simulated intensity maps of the longitudinal electric field at $z$ = 110 nm. The wavelengths are 535 nm (a, b), 625 nm (c, d), 785 nm (e, f), and 855 nm, respectively. The incident polarizations are (a, c, e, g, i) RCP and (b, d, f, h, j) LCP, respectively.

# 5 Chiro-Optical Force Imaging for Enantiomeric Nanostructures

The chiro-optical force imaging for enantiomeric nanostructures is shown in Fig. S8. The results of right-handed gammadion structure are shown in (a-c) and those of left-handed nanostructure in (d-f). The features of the PiFM images for right-handed nanostructures (Figs. S8b and c) are essentially the same as those of Fig. 2 in the main text (i.e., LCP−light induced the strong optical gradient force at the edges of the nanostructure). On the other hand, the opposite trend was observed for the left-handed nanostructures (Figs. S8e and f), i.e., the strong optical gradient force is found at the edges with RCP-light illumination.

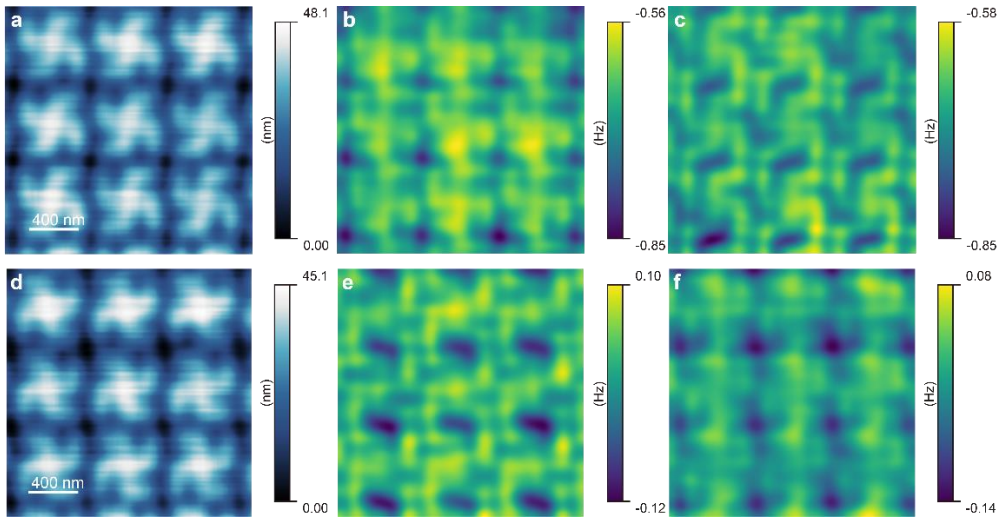

**Figure S8.** PiFM images with the CP-light illumination for right- and left-handed structures. a) Topographic image (AFM) of the right-handed nanostructures. b, c) PiFM image of the right-handed nanostructures under the RCP- and LCP-light illumination, respectively. d) Topographic image (AFM) of the left-handed nanostructure. e, f) PiFM images of the left-handed nanostructures under the RCP- and LCP-light illumination, respectively.

# 6 Imaging with Dissymmetry Factor of Photoinduced Force

To discuss the efficiency of chiro-optical activity, the dissymmetry factor, which is defined as the difference between the physical quantity under LCP and RCP illumination divided by the average of them, is an important parameter. The dissymmetry factor of the PiFM image is illustrated in Fig. S9 ($g_{\Delta f}$ = $-2[\Delta f(f_m)X_L - \Delta f(f_m)X_R]/[\Delta f(f_m)X_L + \Delta f(f_m)X_R]$), and the image shows high $g_{\Delta f}$ values of about 0.24 at the edge positions. We should be careful that $g_{\Delta f}$ also is not completely adequate for the discussion of the nanoscopic chiro-optical effect of the optical gradient force, because of the same reason as the differential PiFM image in Fig. 3 of the main text (the detected signal includes not only optical gradient force but also optical scattering force, and the scattered light is not the same for RCP and LCP illumination: the contribution of scattering force would make the validity of the observed nanoscopic chiro-optical activity worse). In a similar manner to the discussion of the differential PiFM image, $g_{\Delta f}$ obtained under the small cantilever amplitude could be appropriate and useful for quantitative discussion of the local chiro-optical properties.

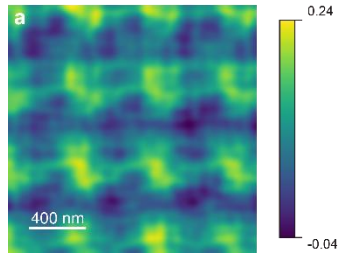

**Figure S9.** The image of the dissymmetry factor of the PiFM image. This image ($g_{\Delta f}$) is calculated from Figs. 2b and c in the main text.